\documentclass[aps,prb,superscriptaddress,letterpaper,amsmath,amssymb,twocolumn,10pt,floatfix]{revtex4-2}

\usepackage{graphicx}
\usepackage[ansinew]{inputenc} 
\usepackage{array} 
\usepackage{xcolor} 
\usepackage{amsxtra}
\usepackage{amstext}
\usepackage{amsthm} 
\usepackage{latexsym}
\usepackage{verbatim} 
\usepackage{bm} 
\usepackage{hyperref} 

\makeatletter
\newcommand\RedeclareMathOperator{%
  \@ifstar{\def\rmo@s{m}\rmo@redeclare}{\def\rmo@s{o}\rmo@redeclare}%
}
\newcommand\rmo@redeclare[2]{%
  \begingroup \escapechar\m@ne\xdef\@gtempa{{\string#1}}\endgroup
  \expandafter\@ifundefined\@gtempa
     {\@latex@error{\noexpand#1undefined}\@ehc}%
     \relax
  \expandafter\rmo@declmathop\rmo@s{#1}{#2}}
\newcommand\rmo@declmathop[3]{%
  \DeclareRobustCommand{#2}{\qopname\newmcodes@#1{#3}}%
}
\@onlypreamble\RedeclareMathOperator
\makeatother

\RedeclareMathOperator{\Re}{Re}
\RedeclareMathOperator{\Im}{Im}

\begin{document}


\title{Non-Hermitian dispersion sign reversal of radiative resonances in two dimensions}

\author{R.~\surname{Binder}}
\affiliation{Wyant College of Optical Sciences, The University of Arizona, Tucson, AZ 85721}
\affiliation{Department of Physics, The University of Arizona, Tucson, AZ 85721}

\author{J.R.~\surname{Schaibley}}
\affiliation{Department of Physics, The University of Arizona, Tucson, AZ 85721}

\author{N.H.~\surname{Kwong}}
\affiliation{Wyant College of Optical Sciences, The University of Arizona, Tucson, AZ 85721}

\date{\today}

\begin{abstract}
In a recent publication [Wurdack et al., Nat. Comm. 14:1026 (2023)], it was shown that in microcavities containing atomically thin semiconductors non-Hermitian quantum mechanics can lead to negative exciton polariton masses. We show that mass-sign reversal  can occur generally in radiative resonances in two dimensions (without cavity) and derive conditions for it (critical dephasing threshold etc.). In monolayer transition-metal dichalcogenides, this phenomenon is not invalidated by the strong electron-hole exchange interaction, which is known to make the exciton massless.
\end{abstract}

\maketitle




Non-Hermitian quantum mechanics
 has recently attracted much interest, both in terms of general physics
 (a recent review article is Ref. \cite{ashida-etal.2020}, a recent textbook Ref.
 \cite{moiseyev.2011})
and also in the area of optics
(review article Ref. \cite{elganainy-etal.2019}).
While the concept of exceptional points
\cite{heiss.00,%
dembowski-etal.01,%
heiss.04}
 is certainly one major driving force behind
this increasing interest
(see for example Refs. \cite{%
choi-etal.10,%
liertzer-etal.12,%
gao-etal.15,%
joshi-galbraith.2018,%
miri-etal.19,%
pan-etal.19,%
sakhdari-etal.19,%
kawabata-etal.19,%
khurgin.2020,%
ohashi-etal.20,%
hanai-littlewood.20,%
ozturk-etal.2021,%
binder-kwong.2021,%
li-etal.2022%
}),
it is only now emerging that non-Hermtian quantum mechanics can lead to other novel phenomena that are not expected on the basis of Hermitian physics.
For example, in Ref. \cite{wurdack-etal.2023} it was
shown that in a microcavity containing a
transition-metal dichalcogenides (TMD) monolayer,
non-Hermitian quantum mechanics can
drastically change the exciton dispersion
\cite{yu-etal.14,wu-etal.15prb,schneider-etal.2019,sauer-etal.2021}
and
 lead to negative exciton polariton
masses. Mass renormalization is a standard concept in many-particle physics; for example,
the coupling of electrons to the lattice vibrations leads to quasiparticles called polarons,
whose effective mass differs from that of the electrons (p. 496 of  \cite{mahan.81}).
 In semiconductor quantum well microcavities, coupling of an exciton resonance to the light field  yields the polariton whose effective mass
is usually much smaller than that of the exciton (e.g. \cite{deng-etal.10}).
But non-Hermitian effects are usually associated with simple line broadenings or lifetime reductions,
not with qualitative effects such as strong mass renormalization or even reversal of the sign of the effective mass.

 We show here that the phenomenon of mass sign reversal, demonstrated in Ref. \cite{wurdack-etal.2023} for polaritons in semiconductor microcavities,
 is not contingent on the cavity;
it applies to any two-dimensional (2D) system with a massive resonance coupled to
the radiation field, or it can generate a finite negative mass if the mass of the resonance is infinite before the coupling.
 We call these electromagnetic modes coupled to the material polarization `2D-layer polaritons' (in the literature they are also sometimes referred to as quantum well polaritons).
 We derive an analytic expression for the critical dephasing at
which the mass diverges and changes sign, and we discuss an analytical
condition for such a sign reversal to be possible (which we find to be strongly affected
by the dielectric environment). While the effective mass is related to the second-order Taylor expansion of the dispersion with respect to the wave vector,
we find that
for non-zero dephasings that are much
smaller than the critical dephasing, a sign reversal of the dispersion of
orders higher than two leads to an energy minimum on a ring at the
edge of the radiative cone. Similar to the predictions for microcavities \cite{wurdack-etal.2023},
the effective mass reversal will affect possible Bose
Einstein condensates (BECs). In the present case of
single layers or quantum wells, we speculate
this would either lead to conical emission or
directional symmetry breaking of the emission (i.e. the BEC would have one more broken-symmetry variable in addition to the phase of the condensate wave function).
We also clarify that the sign reversal is not affected by the
long-range electron-hole exchange interaction
\cite{bayer-etal.2002,yu-etal.14,wu-etal.15prb,qiu-etal.15,steinhoff-etal.2018,deilmann-thygesen.19,sauer-etal.2021}, which is known to
be strong in monolayer TMDs and to
 give
approximately linear exciton dispersions.

We assume the 2D system (e.g. TMD monolayer or thin quantum well) to be in the
x-y plane at $z=z_{0}=0$ and to have a discrete optical resonance described by
the polarization $\mathbf{P}^{3D}=\delta (z-z_{0})\mathbf{P}(\mathbf{q}%
,\omega )$, where $\mathbf{P}(\mathbf{q},\omega )$ is the 2D polarization.\
We assume the layer to be sufficiently thin so that the z-dependence of $%
\mathbf{P}^{3D}$ can be approximated by the $\delta $-function. Maxwell's
propagation equation then reads
\begin{subequations}
\begin{align}
& \left\{ -q^{2}+\frac{\partial ^{2}}{\partial z^{2}}+\epsilon _{b}\frac{%
\omega ^{2}}{c^{2}}\right\} E_{j}(\mathbf{q},z,\omega ) =  \nonumber \\
& - \delta
(z-z_{0})\eta _{v}\left[ \frac{4\pi \omega ^{2}}{c^{2}}P_{j}(\mathbf{q}%
,\omega )-\frac{4\pi }{\epsilon _{b}}q_{j}\mathbf{q}\cdot \mathbf{P}(\mathbf{%
q},\omega )\right]
\end{align}
\end{subequations}
  Here, $j=x,y$ labels the Cartesian component, $q=\sqrt{%
q_{x}^{2}+q_{y}^{2}}$ is the magnitude of the in-plane wavevector, \  $%
\epsilon _{b}$ the background dielectric constant, and $\eta _{v}$ the
multiplicity of equivalent valleys ($\eta _{v}=3$ in TMDs, $\eta _{v}=1$ in
GaAs). The last term in the square bracket stems from the $\triangledown (%
\mathbf{\triangledown }\cdot \mathbf{E})$ term in the propagation equation,
which in turn comes from the $\nabla \times (\nabla \times $ $\mathbf{E)}$
term. In the following, we assume $\mathbf{q}=(q_{x},0)$, so that $E_{x}$ is
the longitudinal (L) and $E_{y}$ the transverse (T) field component. While
the following discussion is valid for any physical realization of the
optical resonance, we use terminology appropriate for our example, which is
the 1s exciton in a direct-gap semiconductor such as a GaAs quantum well
\cite{%
cundiff-etal.92,%
wang-etal.98,%
sieh-etal.99,%
prineas-etal.00,%
meier-etal.00,%
phelps-etal.2011%
}
or semiconducting TMD monolayer
\cite{%
jones-etal.13,%
qiu-etal.13,%
liu-etal.14,%
chernikov-etal.14,%
ye-etal.14,%
yu-etal.14,%
steinhoff-etal.14,%
bellus-etal.15,%
wu-etal.15prb,%
you-etal.15,%
moody-etal.15,%
yu-etal.15,%
hao-etal.16NaturePhysics,%
scuri-etal.18,%
niehues-etal.2018,%
stier-etal.18,%
vantuan-etal.2018,%
mahon-etal.2019,%
selig-etal.2019,%
katsch-etal.2020PRL,%
low-etal.2020,%
yu-etal.2023%
}.
 The polarization components $%
P_{x,y}(\mathbf{q},\omega )=\eta _{v}D_{0}^{\ast }p_{x,y}^{1s}(\mathbf{q}%
,\omega )$ are then a product of the interband dipole matrix element $%
D_{0}=er_{cv}\phi _{1s}(r=0)$ , where $e$ is the electron charge and $er_{cv}
$ the interband dipole matrix element (for a detailed discussion see Ref.
\cite{gu-etal.13}), and the interband coherence of the 1s exciton $%
p_{x,y}^{1s}$, which in linear optical response obeys the equation of motion
\begin{equation}
i\hbar \frac{\partial }{\partial t}p_{j}^{1s}(\mathbf{q},t)=\left(
\varepsilon _{q}^{1s}-i\gamma _{D}\right) p_{j}^{1s}(\mathbf{q}%
,t)-D_{0}E_{j}(\mathbf{q},z_{0},t)
\end{equation}
(where $\gamma _{D}$ is the dephasing), or, after Fourier transformation,%
\begin{equation}
0=\left( \varepsilon _{q}^{1s}-i\gamma _{D}-\hbar \omega \right) p_{j}^{1s}(%
\mathbf{q},\omega )-D_{0}E_{j}(\mathbf{q},z_{0},\omega )
\end{equation}

  We use a massive (parabolic) dispersion $\varepsilon _{q_{x}}^{1s}=%
\frac{\hbar ^{2}q_{x}^{2}}{2m_{x}}+\varepsilon _{0}^{1s}$ with the exciton
mass $m_{x}$.
Solving the Maxwell equation with a transfer matrix method
\cite{sipe.87,khitrova-etal.99}
 and assuming only
outgoing waves (no light field incident on the layer), one obtains the
dispersion relation for 2D systems in a form that is by now well known (see,
for example,
\cite{epstein-etal.2020,klein-etal.2019}  and references therein),
namely
\begin{equation}
\epsilon _{b}/k_{z}-i2\pi \chi (\mathbf{q},\omega )=0
\end{equation}%
for the longitudinal waves, and
\begin{equation}
k_{z}-i2\pi (\omega ^{2}/c^{2})\chi (\mathbf{q},\omega )=0
\end{equation}%
for transverse waves, with the 2D susceptibility
\begin{equation}
\chi (\mathbf{q},\omega )=\frac{\eta _{v}|D_{0}|^{2}}{\varepsilon
_{q}^{1s}-i\gamma _{D}-\hbar \omega }
\end{equation}%
and $k_{z}=\sqrt{\epsilon _{b}(\omega ^{2}/c^{2})-q_{x}^{2}}$. The
dispersion relation is an implicit equation for $\omega _{L}(q_{x})$ and $%
\omega _{T}(q_{x})$, respectively, and   we choose $q_{x}$ real-valued
and $\omega $ complex-valued.

\begin{figure}
	\centering
	\includegraphics[width=3.7in]{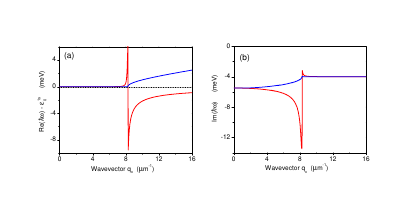}
	\caption{
		(Color online.)
		Real (a) and imaginary (b) part of the dispersion relation of longitudinal (blue) and transverse (red) 2D-layer polaritons. The dephasing is $\gamma_D = 4$ meV.
The edge of the radiative cone is approximately at 8.23 $\mu$m$^{-1}$. The real (imaginary) part of $\hbar \omega_L$ is discontinuous (continuous) at the edge of the radiative cone;
the real and imaginary part of $\hbar \omega_T$ are discontinuous.
	}
\label{fig:sr3174LandT}
\end{figure}

Figure \ref{fig:sr3174LandT} shows an example for the longitudinal and transverse
waves using parameters typical for monolayer MoSe$_{2}$, with
$\gamma _{D}=4$
meV, inside and outside the radiative cone, which is at $k_{c}(\omega )=%
\sqrt{\epsilon _{b}}(\omega /c)$. These results look similar to those shown
in Ref. \cite{tassone-etal.90}, except for the T-branch inside the cone (we
believe an effective-mass approximation, which is not valid close to the
edge of the radiative cone, was used in that reference). Moreover,
\cite{tassone-etal.90} finds that both L and T modes inside the radiative cone
have a curvature (or effective mass) similar to the free exciton mass, $m_{x}
$. In Ref. \cite{tassone-etal.90},
 the dephasing was zero or negligible. In our
results, the total decay of the eigenmode includes the radiative decay and
the dephasing, and therefore the curves for the imaginary parts of $\omega
(q_{x})$ are approximately shifted by \ $\gamma _{D}$.

We will now show that the inclusion of the dephasing \ $\gamma _{D}$ has
important consequences for the dispersion of the longitudinal 2D-layer
dispersion $\omega _{L}(q_{x})$, highlighting the effects of non-Hermitian
quantum mechanics, which, as mentioned, had been pointed out in
\cite{wurdack-etal.2023} for the case of microcavities. We will analyze
the behavior of the L-mode inside the radiative cone of a
2D system with a massive radiative resonance. We suppress from now
on the subscript L.

We expand the dispersion relation to lowest order in $q_{x}$ (which is $%
q_{x}^{2}$). We write the dispersion $\omega (q_{x})=\omega _{0}+\Delta
\omega (q_{x})$ and keep only terms linear in $\Delta \omega (q_{x})$. For
the solution at $q_{x}=0$, we find
\begin{equation}
\hbar \omega _{0}=\frac{\varepsilon _{0}^{1s}-i\gamma _{D}}{1+i\gamma
_{R}/\varepsilon _{0}^{1s}}
\end{equation}
where $\gamma _{R}=\frac{2\pi }{\sqrt{\epsilon _{b}}\hbar c}\eta
_{v}|D_{0}|^{2}\varepsilon _{0}^{1s}$ is the radiative decay at $q_{x}=0$.
This equation contains both the radiative shift of the resonance and its
decay. The complex-valued dispersion at non-zero $q_{x}$ reads (to order $%
q_{x}^{2}$)
\begin{equation}
\hbar \Delta \omega (q_{x})=\frac{1}{1+i\gamma _{R}/\varepsilon _{0}^{1s}}%
\frac{\hbar ^{2}q_{x}^{2}}{2m_{x}}+i\frac{\hbar ^{2}c^{2}\gamma _{R}q_{x}^{2}%
}{2\epsilon _{b}\varepsilon _{0}^{1s}(\varepsilon _{0}^{1s}-i\gamma _{D})}
\end{equation}%
For the real part, we therefore obtain
\begin{equation}
{\rm Re}\hbar \Delta \omega (q_{x})=\frac{1}{1+(\gamma _{R}/\varepsilon
_{0}^{1s})^{2}}\frac{\hbar ^{2}q_{x}^{2}}{2m_{x}}-\frac{\hbar
^{2}c^{2}\gamma _{R}}{2\epsilon _{b}\varepsilon _{0}^{1s}}\frac{\gamma _{D}}{%
(\varepsilon _{0}^{1s})^{2}+\gamma _{D}^{2}}q_{x}^{2}
\label{Eq:ReDeltaOmegaqx}
\end{equation}%
We can use this equation to define an effective radiative exction mass $%
m_{Rad}$ via
\begin{equation}
{\rm Re}\hbar \Delta \omega (q_{x})=\frac{\hbar ^{2}q_{x}^{2}}{2m_{Rad}}
\end{equation}

\begin{figure}
	\centering
	\includegraphics[width=3.7in]{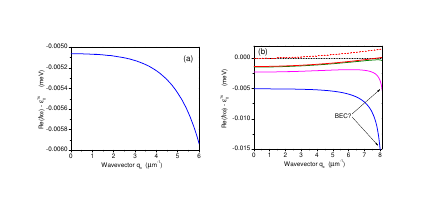}
	\caption{
		(Color online.)
		(a) Same as Fig. \protect\ref{fig:sr3174LandT}(a) for longitudinal mode, but only showing inside the radiative cone. (b) Same as (a) for blue curve, plus dispersion for $\gamma_D=0$ (red curve), 0.1 meV (green), 1 meV (magenta). The red dashed curve shows the exciton dispersion.
	}
\label{fig:sr3174InsideCone}
\end{figure}

The first term in Eq. (\ref{Eq:ReDeltaOmegaqx}) comes from the exciton
dispersion and contains a radiative correction to the effective mass. This
correction is present even without dephasing. However, in our numerical
example, this correction is negligible.

The second term is another radiative correction to the dispersion, but one
that is present only if the dephasing is non-zero. Its size depends
therefore on the value of $\gamma _{D}$. To get an estimate of the
importance of this term, we define a critical dephasing $\gamma _{D}^{c}$
such that the curvature vanishes, ${Re}\hbar \Delta \omega
(q_{x};\gamma _{D}^{c})=0$, or in other words, the radiative 2D-layer
polariton mass diverges. We find two critical dephasings,
\begin{equation}
\gamma _{D}^{c(+/-)}=\frac{1}{2}\varepsilon _{Rad}\left( 1\pm \sqrt{%
1-4\left( \varepsilon _{0}^{1s}/\varepsilon _{Rad}\right) ^{2}}\right)
\end{equation}
  with $\varepsilon _{Rad}=\gamma _{R}\left[ 1+(\gamma
_{R}/\varepsilon _{0}^{1s})^{2}\right] m_{x}c^{2}/(\epsilon _{b}\varepsilon
_{0}^{1s})$. As we will show below, the smaller critical dephasing $\gamma
_{D}^{c(-)}$ is more relevant than $\gamma _{D}^{c(+)}$, but both give us a
complete picture of the evolution of the radiative mass as the dephasing is
varied from zero to infinity. The requirement that the square root in this
equation is real-valued gives us the following condition for the possibility
of a sign reversal of the effective radiative mass:
\begin{equation}
2\epsilon _{b}\left( \varepsilon _{0}^{1s}\right) ^{2}\leq m_{x}c^{2}\gamma
_{R}\left[ 1+(\gamma _{R}/\varepsilon _{0}^{1s})^{2}\right]
\end{equation}%
which can also be written as
\begin{equation}
\epsilon _{b}^{3/2}\varepsilon _{0}^{1s}\leq \widetilde{m}_{x}\eta _{v}|%
\widetilde{D}_{0}|^{2}\widehat{\varepsilon }\left[ 1+(\gamma
_{R}/\varepsilon _{0}^{1s})^{2}\right]
\label{Eq:second-less-than-condition}
\end{equation}%
where we define $m_{x}=\widetilde{m}_{x}m_{0}$ ($m_{0}$ being the electron
mass in vacuum), and $|D_{0}|^{2}=|\widetilde{D}_{0}|^{2}10^{-9}$eVcm (since
in TMDs and GaAs $|D_{0}|^{2}$ is approximately $10^{-9}$eVcm), and $%
\widehat{\varepsilon }=\pi m_{0}c^{2}10^{-9} {\rm eVcm}/\hbar c\approx 80$eV. The
relatively small value of $\widehat{\varepsilon }$ shows that the
possibility of the mass sign reversal depends sensitively on the exact
numbers, such as the dielectric constant of the background material and the
effective exciton mass. For example, for TMDs we use $|D_{0}|^{2}=0.96\times
10^{-9}$ eVcm, $\widetilde{m}_{x}=1.67$, $\eta _{v}=3$, and $\epsilon _{b}=1$
(for free-standing monolayers), and $\varepsilon _{0}^{1s}=1.6$ eV. Then the
left-hand side (LHS) of Eq. (\ref{Eq:second-less-than-condition}) is $1.6$ eV and the right-hand side (RHS)
about $384$eV (where we neglect the correction from the square bracket). To
estimate a GaAs QW embedded in GaAlAs, we take $|D_{0}|^{2}=0.46\times
10^{-9}$ eVcm, $\widetilde{m}_{x}=0.2$, $\eta _{v}=1$, and $\epsilon _{b}=16$%
. Now the LHS increases to 96eV (mostly because of the dielectric
environment described by the factor $\epsilon _{b}^{3/2}$), while the RHS
reduces to 9eV. Hence the condition for mass sign reversion is not
fulfilled. If, however, one could us a free-standing QW membrane (if bending and wrinkling can be avoided
\cite{cendula-etal.09}), the condition would be fulfilled. Similarly, our numerical values for $%
\gamma _{R}$, which are 1.5 meV in TMDs and 0.057 meV in GaAs, show that even the
large factor $m_{x}c^{2}$ is not sufficient for the condition to be
fulfilled in GaAs in the presence of a strong dielectric environment. We
note that one could have a dielectric environment with $\epsilon _{b}\approx
0$, in so-called epsilon-near-zero (ENZ) materials (e.g. \cite{liberal-engheta.2017}). In such an environment, the mass sign
reversal would work even for arbitrarily small $m_{x}$ and $\gamma _{R}$.

\begin{figure}
	\centering
	\includegraphics[width=3.7in]{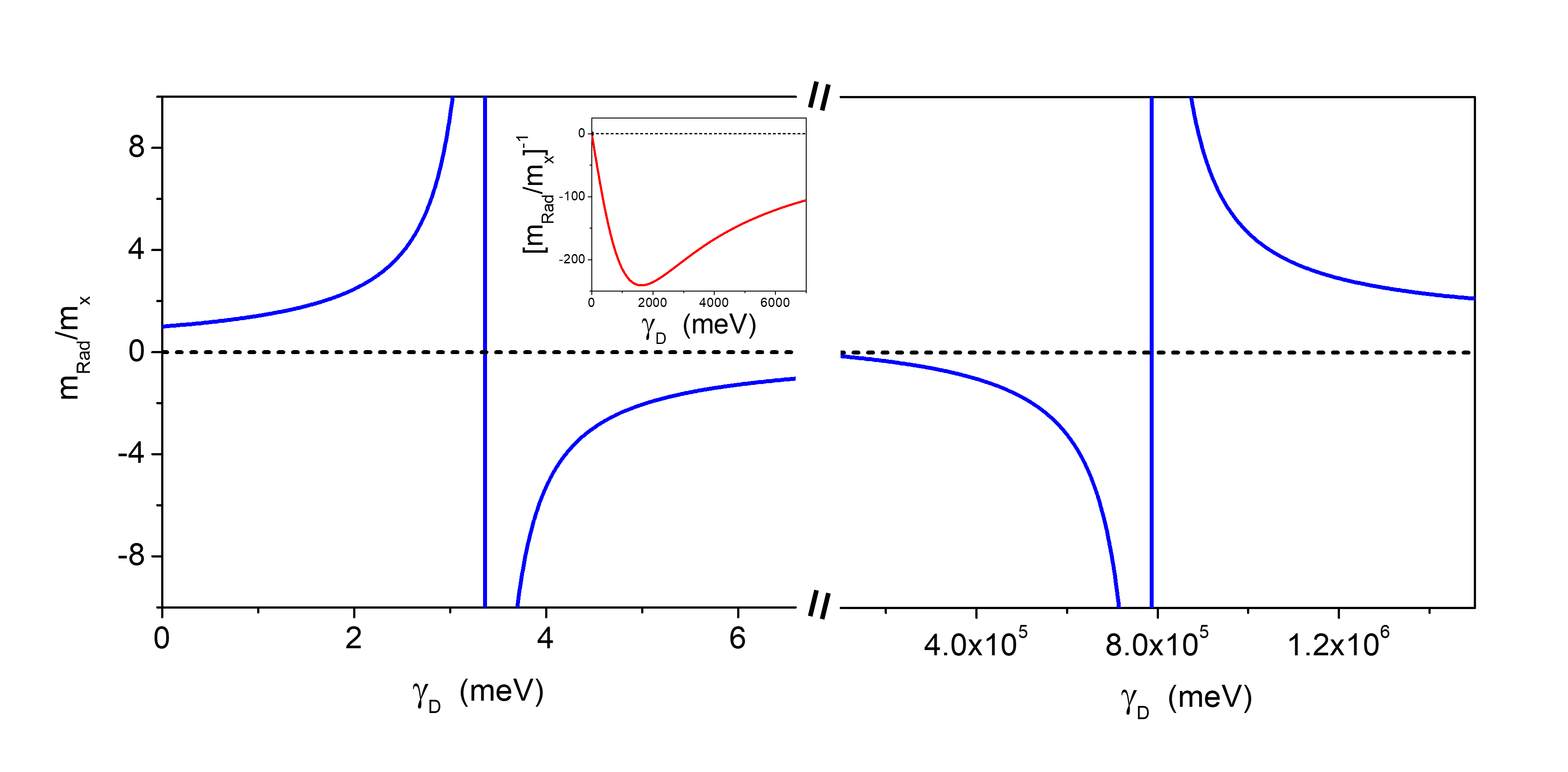}
	\caption{
		(Color online.)
		Radiative mass in units of exciton mass vs. dephasing. The inset shows the inverse radiative mass in the region of the extremum.
	}
\label{fig:sr3170-masses}
\end{figure}

  In our numerical example, we find $\gamma _{R}\ll \varepsilon
_{0}^{1s}$ \ for TMDs and GaAs, and for TMDs we also find $\gamma _{D}\ll
\varepsilon _{0}^{1s}$. In the latter case, the expression for the radiative
exciton mass simplifies to
\begin{equation}
m_{Rad}\approx \frac{m_{x}}{1-\gamma _{D}/\gamma _{D}^{c}}
\end{equation}
with
\begin{equation}
\gamma _{D}^{c}\approx \frac{\hbar ^{2}}{2m_{x}}\frac{2\epsilon _{b}\left(
\varepsilon _{0}^{1s}\right) ^{2}}{\hbar ^{2}c^{2}\gamma _{R}}
\end{equation}
 Note that this approximate formula, when applied to TMDs, gives a
reasonable estimate of $\gamma _{D}^{c}=3.3$ meV, while for GaAs this gives a
value much larger than $\varepsilon _{0}^{1s}$ and therefore contradicts the
assumption for the approximate formula.

We also find that, in the case of infinite exciton mass $m_x = \infty$ (localized excitons), the non-Hermitian coupling creates a finite mass. In
this case, the first term in Eq. (\ref{Eq:ReDeltaOmegaqx}) is zero, and, to order $q_x^2$, the second term creates an effective negative mass,
for any non-zero dephasing (i.e. $\gamma_D^c = 0$).

Continuing the discussion of Fig. \ref{fig:sr3174LandT} for monolayer  MoSe$_{2}$,
we see in Fig. \ref{fig:sr3174InsideCone} (a) that for $\gamma _{D}=4$ meV the
curvature is negative throughout the entire radiative cone, implying a
negative radiative mass $m_{Rad}$. To verify that 4 meV is indeed above the
critical value, we show in Fig. \ref{fig:sr3170-masses} the variation of the
effective mass as a function of dephasing. We see that $\gamma
_{D}^{c(-)}=3.36$ meV, indeed smaller than 4 meV. As the dephasing goes from
zero to infinity, the first singularity of the radiative mass is at $\gamma
_{D}^{c(-)}$. It then has an extremum at about 1.56eV, and then another
singularity at $\gamma _{D}^{c(+)}$, which for practical purposes does not
seem relevant.

  It is, however, not only the effective mass, that changes sign as
the dephasing increases. Figure \ref{fig:sr3174InsideCone} (b) shows that for
(numerically) abritrarily small $\gamma _{D}$ the dispersion relation `rolls
over' close to the edge of the radiative cone. This implies that in a
higher-order Taylor expansion of $\Delta \omega (q_{x})$ we probably have
different critical dephasings for each order in $q_{x}^{2}$. Interestingly, the
groundstate of the longitudinal 2D-layer polariton is, for many values of $%
\gamma _{D}$, at the edge of the radiative cone. This brings up the question
of possible excitonic Bose-Einstein condensates (BECs) \cite{morita-etal.2022}. If the BEC forms at the edge of
the radiative cone, then there are two possible scenarios: (i) the BEC
emission would be conical (if the entire ring contains the BEC),
or there is a second spontaneous symmetry breaking (in addition
to the U(1) phase symmetry that is related to the BEC even at zero wave
vector) which would choose a certain point on the ring of the radiative cone
and thus lead to  directed emission with arbitrary direction. However, whether or not a BEC
can occur in this system still needs further research, including the issue of opposite
temperature requirements (BEC benefits from low temperatures, large dephasing from high temperatures).

\begin{figure}
	\centering
	\includegraphics[width=3.7in]{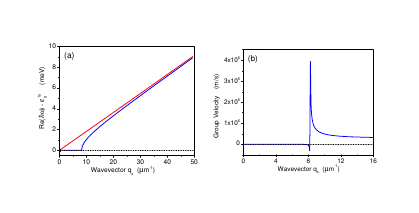}
	\caption{
		(Color online.)
		(a) Blue curve same as blue curve in Fig. \protect\ref{fig:sr3174LandT}. The redline shows $\varepsilon _{q_{x},L}^{1s,static}$.
(b) Group velocity corresponding to blue curve in (a).
	}
\label{fig:sr3174ExchangeVg}
\end{figure}

We finally discuss the issue of long-range electron-hole (e-h) exchange interaction, which in TMD
monolayers is known to be very strong, and which is related to a linear (in $%
q_{x}$) dispersion
(e.g.
\cite{%
qiu-etal.13,%
yu-etal.14,%
wu-etal.15prb,%
qiu-etal.15,%
sauer-etal.2021,%
kwong-etal.2021}).
 In other words, it has been found to make the
exciton massless. \ This issue has been in principle already addressed in
Ref. \cite{tassone-etal.90}, where it was pointed out that the self-consistent solution
of the radiation field and the material response includes the effects of
long-range e-h exchange interaction. If the latter is treated with a static Coulomb
interaction, then that treatment is only valid at wavevectors much larger
than the radiative cone, for example on a wavevector scale that is relevant for the binding of two excitons into a biexciton
\cite{kwong-etal.2021}.
However, some publications extend the linear
dispersion obtained from a static Coulomb interaction explicitly down to wavevectors inside the cone, where the assumption of
static Coulomb interactions and the neglect of radiation retardation effects is not valid \cite{tassone-etal.90}.

 To clarify this point,
we show in Fig. \ref{fig:sr3174ExchangeVg} (a) the dispersion on a wavevector
scale six times larger than the radiative cone. We also plot the result from the static
calculation
 (cf. \cite{yu-etal.14,wu-etal.15prb,steinhoff-etal.2018,schneider-etal.2019,kwong-etal.2021}),
\begin{equation}
\varepsilon _{q_{x},L/T}^{1s,static}=\frac{\hbar ^{2}q_{x}^{2}}{2m_{x}}%
+\varepsilon _{0}^{1s}+J^{exch}_{L/T}(q_{x})
\end{equation}%
with
$J^{exch}_{L/T}(q_{x})=J^{intra}(q_{x}) \pm J^{inter}(q_{x})$. The exchange contributions for L  and T are
$J^{exch}_{L}(q_{x})= \eta _{v} 2\pi e^{2} |D_{0}|^{2}q_{x} / \epsilon _{b} $
 and  $J^{exch}_{T}(q_{x})=0$, respectively.
For L it includes the sum of intervalley and intravally e-h exchange (these contributions cancel each other for T).
Here, we omit $q_{x}$%
-dependent screening, as would be included in the the Rytova-Keldysh
interaction and improved versions thereof \cite{vantuan-etal.2018}. We see that, at wavevectors considerably larger than the radiative cone
(but still small enough for the $q_x^2$-term to be negligible)
the linear dispersion from the e-h exchange is included in $\omega
_{L}(q_{x})$, as was noted in Ref. \cite{tassone-etal.90}. Importantly, we find that
the effect of mass sign reversal (and more generally dispersion sign
reversal) is not affected by the long-range e-h exchange interaction, as it is a
phenomenon limited to the inside of the radiative cone. We also show in Fig.
\ref{fig:sr3174ExchangeVg} (b) the group velocity of the L-waves, which is
relatively large close to the radiative cone, and in our example
negative inside the cone.


In summary, we have shown that, as a consequence of non-Hermitian coupling, 2D-layer polaritons can exhibit mass-sign reversal similar to microcavity polaritons and have derived an analytic expression that sets conditions on the mass sign reversal.
We noted that even without the mass sign reversal, the dispersion can roll over at the radiative cone and discussed hypothetical scenarious for BECs with ground states on a ring. Future research is needed into the condensation behaviour in these systems. 
Our findings may also be relevant for TMD-based lasers
(e.g. \cite{ye-etal.15,%
wu-etal.15Nature,%
li-CZNing-etal.17}).
%
 Finally, we clarified that the long-range e-h exchange interaction does not affect the sign reversal  inside the radiative cone.

RB acknowledges
financial support from the US National Science Foundation (NSF) under grant number DMR 1839570,
and the use of High Performance Computing (HPC) resources supported by the University of Arizona.
JRS acknowledges funding from AFOSR Grant number: FA9550-20-1-0217.




\begin{thebibliography}{10}

\bibitem{ashida-etal.2020}
Y. Ashida, Z. Gong, and M. Ueda, Advances in Physics {\bf 69},  249   (2020).

\bibitem{moiseyev.2011}
N. Moiseyev, {\em Non-Hermitian Quantum Mechanics} (Cambridge, Cambridge,
  2011).

\bibitem{elganainy-etal.2019}
{Ramy El-Ganainy and Mercedeh Khajavikhan and Demetrios N. Christodoulides and
  Sahin K. Ozdemir}, Communications Physics {\bf 2},  37  (2019).

\bibitem{heiss.00}
W.~D. Heiss, Physical Rev. E {\bf 61},  929  (2000).

\bibitem{dembowski-etal.01}
C. Dembowski, H.-D. Graf, H.~L. Harney, A. Heine, W.~D. Heiss, H. Rehfeld, and
  A. Richter, Phys. Rev. Lett. {\bf 86},  787   (2001).

\bibitem{heiss.04}
W.~D. Heiss, Journal of Physics: Mathematical and General {\bf 37},  2455
  (2004).

\bibitem{choi-etal.10}
Youngwoon Choi, Sungsam Kang, Sooin Lim, Wookrae Kim, Jung-Ryul Kim, Jai-Hyung
  Lee, and Kyungwon An, Phys. Rev. Lett. {\bf 104},  153601  (2010).

\bibitem{liertzer-etal.12}
M. Liertzer, L. Ge, A. Cerjan, A.~D. Stone, H.~E. Tureci, and S. Rotter, Phys.
  Rev. Lett. {\bf 108},  173901   (2012).

\bibitem{gao-etal.15}
T. Gao, E. Estrecho1, K.~Y. Bliokh, T.~C.~H. Liew, M.~D. Fraser, S. Brodbeck,
  M. Kamp, C. Schneider, S. Hoefling, Y. Yamamoto, F. Nori, Y.~S. Kivshar,
  A.~G. Truscott, R.~G. Dall, and E.~A. Ostrovskay, Nature {\bf 526},  554
  (2015).

\bibitem{joshi-galbraith.2018}
Sharad Joshi and Ian Galbraith, Phys. Rev. A {\bf 98},  042117  (2018).

\bibitem{miri-etal.19}
M.~A. Miri and A. Alu, Science Magazine {\bf 363},  42  (2019).

\bibitem{pan-etal.19}
L. Pan, S. Chen, and X. Cui, Physical Rev. A {\bf 99},  011601  (2019).

\bibitem{sakhdari-etal.19}
M. Sakhdari, M. Hajizadegan, Q. Zhong, D.~N. Christodoulides, R. El-Ganainy,
  and P.~Y. Chen, Phys. Rev. Lett. {\bf 123},  193901  (2019).

\bibitem{kawabata-etal.19}
K. Kawabata, T. Bessho, and M. Sato, Phys. Rev. Lett. {\bf 123},  066405
  (2019).

\bibitem{khurgin.2020}
J. Khurgin, Optica {\bf 7},  1015   (2020).

\bibitem{ohashi-etal.20}
T. Ohashi, S. Kobayashi, and Y. Kawaguchi, Physical Rev. A {\bf 101},  013625
  (2020).

\bibitem{hanai-littlewood.20}
R. Hanai and P. Littlewood, Phys. Rev. Research {\bf 2},  033018  (2020).

\bibitem{ozturk-etal.2021}
Fahri~Emre Ozturk, Tim Lappe, Goran Hellmann, Julian Schmitt, Jan Klaers, Frank
  Vewinger, Johann Kroha, and Martin Weitz, Science {\bf 372},  88  (2021).

\bibitem{binder-kwong.2021}
R. Binder and N.H. Kwong, Phys. Rev. B {\bf 103},  085304  (2021).

\bibitem{li-etal.2022}
Yao Li, Xuekai Ma, Zaharias Hatzopoulos, Pavlos~G. Savvidis, Stefan Schumacher,
  and Tingge Gao, ACS Photonics {\bf 9},    .

\bibitem{wurdack-etal.2023}
M. Wurdack, T. Yun, M. Katzer, A.~G. Truscott, A. Knorr, M. Selig, E.~A.
  Ostrovskaya, and E. Estrecho, Nat. Comm. {\bf 14},  1026  (2023).

\bibitem{yu-etal.14}
H. Yu, G. Liu, P. Gong, X. Xu, and W. Yao, Nature Communications {\bf 5},  3876
   (2014).

\bibitem{wu-etal.15prb}
F. Wu, F. Qu, and A.~H. MacDonald, Phys. Rev. B {\bf 91},  075310   (2015).

\bibitem{schneider-etal.2019}
L.~M. Schneider, S.~S. Esdaille, D.~A. Rhodes, K. Barmak, J.~C. Hone, and A.
  Rahimi-Iman, Opt. Exp. {\bf 27},  37131  (2019).

\bibitem{sauer-etal.2021}
Mikkel~Ohm Sauer, Carl Emil~M\o{}rch Nielsen, Lars Merring-Mikkelsen, and
  Thomas~Garm Pedersen, Phys. Rev. B {\bf 103},  205404  (2021).

\bibitem{mahan.81}
G.~D. Mahan, {\em Many-Particle Physics} (Plenum Press, New York, 1981).

\bibitem{deng-etal.10}
H. Deng, H. Haug, and Y. Yamamoto, Rev. Mod. Phys. {\bf 82},  1489  (2010).

\bibitem{bayer-etal.2002}
M. Bayer, G. Ortner, O. Stern, A. Kuther, A.~A. Gorbunov, A. Forchel, P.
  Hawrylak, S. Fafard, K. Hinzer, T.~L. Reinecke, S.~N. Walck, J.~P.
  Reithmaier, F. Klopf, and F. Sch\"afer, Phys. Rev. B {\bf 65},  195315
  (2002).

\bibitem{qiu-etal.15}
D.Y. Qiu, T. Cao, and S.G. Louie, Phys. Rev. Lett. {\bf 115},  176801   (2015).

\bibitem{steinhoff-etal.2018}
Alexander Steinhoff, Matthias Florian, Akshay Singh, Kha Tran, Mirco Kolarczik,
  Sophia Helmrich, Alexander~W. Achtstein, Ulrike Woggon, Nina Owschimikow,
  Frank Jahnke, and Xiaoqin Li, Nature Phys. {\bf 14},  1199   (2018).

\bibitem{deilmann-thygesen.19}
T. Deilmann and K.~S. Thygesen, 2D Materials {\bf 6},  035003  (2019).

\bibitem{cundiff-etal.92}
S.~T. Cundiff, H. Wang, and D.~G. Steel, Phys. Rev. B {\bf 46},  7248  (1992).

\bibitem{wang-etal.98}
H. Wang, H.~Q. Hou, and B.~E. Hammons, Phys. Rev. Lett. {\bf 81},  3255
  (1998).

\bibitem{sieh-etal.99}
C. Sieh, T. Meier, F. Jahnke, A. Knorr, S.~W. Koch, P. Brick, M. H{\"{u}}bner,
  C. Ell, J. Prineas, G. Khitrova, and H.M. Gibbs, Phys. Rev. Lett. {\bf 82},
  3112  (1999).

\bibitem{prineas-etal.00}
J.~P. Prineas, C. Ell, E.S. Lee, G. Khitrova, H.~M. Gibbs, and S.~W. Koch,
  Phys. Rev. B {\bf 61},  13863   (2000).

\bibitem{meier-etal.00}
T. Meier, S.~W. Koch, M. Phillips, and H. Wang, Phys. Rev. B {\bf 62},  12605
  (2000).

\bibitem{phelps-etal.2011}
Carey Phelps, John Prineas, and Hailin Wang, Phys. Rev. B {\bf 83},  153302
  (2011).

\bibitem{jones-etal.13}
A.~M. Jones, H. Yu, N.~J. Ghimire, S. Wu, F. Aivazian, J.~S. Ross, B. Zhao, J.
  Yan, D.~G. Mandrus, D. Xiao, W. Yao, and X. Xu, Nature Nanotechnology {\bf
  8},  634  (2013).

\bibitem{qiu-etal.13}
D.~Y. Qiu, F.~H. da~Jornada, and S.~G. Louie, Phys. Rev. Lett. {\bf 111},
  216805   (2013).

\bibitem{liu-etal.14}
{X. Liu and T. Galfsky and Zh. Sun and F. Xia and E. Lin and Y. Lee and S.
  Kena-Cohen and V. Menon}, Nature Photonics {\bf 9},  30   (2014).

\bibitem{chernikov-etal.14}
A. Chernikov, T.~C. Berkelbach, H.~M. Hill, A. Rigosi, Y. Li, O.~B. Aslan,
  D.~R. Reichman, M.~S. Hybertsen, and T.~F. Heinz, Phys. Rev. Lett. {\bf 113},
   076802   (2014).

\bibitem{ye-etal.14}
Z. Ye, T. Cao, K O'Brien, H. Zhu, X. Yin, Y. Wang, S. Louie, and X. Zhang,
  Nature {\bf 513},  214   (2014).

\bibitem{steinhoff-etal.14}
A. Steinhoff, M. Rosner, F. Jahnke, T.~O. Wehling, and C. Gies, Nano Lett. {\bf
  14},  3743  (2014).

\bibitem{bellus-etal.15}
M.~Z. Bellus, F. Ceballos, H.~Y. Chiu, and H. Zhao, ACS Nano {\bf 9},  6459
  (2015).

\bibitem{you-etal.15}
Y. You, X. Zhang, T.~C. Berkelbach, M.~S. Hybertsen, D.~R. Reichman, and T.~F.
  Heinz, Nature Physics {\bf 11},  477  (2015).

\bibitem{moody-etal.15}
G. Moody, C.K. Dass, K. Hao, C. Chen, L. Li, A. Singh, K. Tran, G. Clark, X.
  Xu, G. Bergh{\"{a}}user, E. Malic, A. Knorr, and X. Li, Nature Communications
  {\bf 6},    (2015).

\bibitem{yu-etal.15}
H. Yu, X. Cui, X. Xu, and W. Yao, National Science Review {\bf 2},  57  (2015).

\bibitem{hao-etal.16NaturePhysics}
K. Hao, G. Moody, F. Wu, C.~K. Dass, L Xu., C.~H. Chen, L. Sun, M.~Y. Li, L.~J.
  Li, A.~H. MacDonald, and X. Li, Nature Physics {\bf 12},  677  (2016).

\bibitem{scuri-etal.18}
G. Scuri, Y. Zhou, A.~A. High, D.~S. Wild, C. Shu, K.~De Greve, L.~A. Jauregui,
  T. Taniguchi, K. Watanabe, P. Kim, M.~D. Lukin, and H. Park, Phys. Rev. Lett.
  {\bf 120},  037402   (2018).

\bibitem{niehues-etal.2018}
Iris Niehues, Robert Schmidt, Matthias Druppel, Philipp Marauhn, Dominik
  Christiansen, Malte Selig, Gunnar Berghauser, Daniel Wigger, Robert
  Schneider, Lisa Braasch, Rouven Koch, Andres Castellanos-Gomez, Tilmann Kuhn,
  Andreas Knorr, Ermin Malic, Michael Rohlfing, Steffen~Michaelis
  de~Vasconcellos, and Rudolf Bratschitsch, Nano Lett. {\bf 18},  1751
  (2018).

\bibitem{stier-etal.18}
A.~V. Stier, N.~P. Wilson, K.~A. Velizhanin, J. Kono, X. Xu, and S.~A. Crooker,
  Phys. Rev. Lett. {\bf 120},  057405   (2018).

\bibitem{vantuan-etal.2018}
Dinh Van~Tuan, Min Yang, and Hanan Dery, Phys. Rev. B {\bf 98},  125308
  (2018).

\bibitem{mahon-etal.2019}
P.T. Mahon, R.A. Muniz, and J.E. Sipe, Phys. Rev. B {\bf 99},  235140  (2019).

\bibitem{selig-etal.2019}
Malte Selig, Florian Katsch, Robert Schmidt, Steffen Michaelis~de Vasconcellos,
  Rudolf Bratschitsch, Ermin Malic, and Andreas Knorr, Phys. Rev. Research {\bf
  1},  022007  (2019).

\bibitem{katsch-etal.2020PRL}
Florian Katsch, Malte Selig, and Andreas Knorr, Phys. Rev. Lett. {\bf 124},
  257402  (2020).

\bibitem{low-etal.2020}
A. Chaves, J.~G. Azadani, Hussain Alsalman, D.~R. da~Costa, R. Frisenda, A.~J.
  Chaves, Seung~Hyun Song, Daowei~He Y.~D.~Kim, Jiadong Zhou, A.
  Castellanos-Gomez, F.~M. Peeters, Zheng Liu, C.~L. Hinkle, Sang-Hyun Oh,
  Peide~D. Ye, Steven~J. Koester, Young~Hee Lee, Ph. Avouris, Xinran Wang, and
  Tony Low, npj 2D Materials and Applications {\bf 4},  20  (2020).

\bibitem{yu-etal.2023}
Yueyang Yu, Chuan Ding, Rolf Binder, Stefan Schumacher, and Cun-Zheng Ning, ACS
  Nano {\bf 17},  4230   (2031).

\bibitem{gu-etal.13}
B. Gu, N.H. Kwong, and R. Binder, Phys. Rev. B {\bf 87},  125301   (2013).

\bibitem{sipe.87}
J. Sipe, J. Opt. Soc. B {\bf 4},  481   (1987).

\bibitem{khitrova-etal.99}
G. Khitrova, H.~M. Gibbs, F. Jahnke, M. Kira, and S.~W. Koch, Reviews of Modern
  Physics {\bf 71},  1591  (1999).

\bibitem{epstein-etal.2020}
Itai Epstein, Andre~J Chaves, Daniel~A Rhodes, Bettina Frank, Kenji Watanabe,
  Takashi Taniguchi, Harald Giessen, James~C Hone, Nuno M~R Peres, , and Frank
  H~L Koppens, 2D Materials {\bf 7},  035031  (2020).

\bibitem{klein-etal.2019}
Matthew Klein, Bekele~H. Badada, Rolf Binder, Adam Alfrey, Max McKie,
  Michael~R. Koehler, David~G. Mandrus, Takashi Taniguchi, Kenji Watanabe,
  Brian~J. LeRoy, and John~R. Schaibley, Nat. Commun. {\bf 10},  3264  (2019).

\bibitem{tassone-etal.90}
F. Tassone, F. Bassani, and L.C. Andreani, Nuovo Cimento D {\bf 12},  1673
  (1990).

\bibitem{cendula-etal.09}
P. Cendula, S. Kiravittaya, Y.F. Mei, Ch. Deneke, and O.G. Schmidt, Phys. Rev.
  B {\bf 79},  085429  (2009).

\bibitem{liberal-engheta.2017}
{Inigo Liberal and Nader Engheta}, Nat. Photonics {\bf 11},  149   (2017).

\bibitem{morita-etal.2022}
Y. Morita, K. Yoshioka, and M. Kuwata-Gonokami, Nat. Comm. {\bf 13},  5388
  (2022).

\bibitem{kwong-etal.2021}
N.~H. Kwong, J.~R. Schaibley, and R. Binder, Phys. Rev. B {\bf 104},  245434
  (2021).

\bibitem{ye-etal.15}
Y. Ye, Z.~J. Wong, X. Lu, X. Ni, H. Zhu, X. Chen, Y. Wang, and X. Zhang, Nature
  Photonics {\bf 9},  733  (2015).

\bibitem{wu-etal.15Nature}
S. Wu, S. Buckley, J.~R. Schaibley, L. Fengand, J. Yan, D.~G. Mandrus, F.
  Hatami, W. Yao, J. Vuckovi{\'{c}}, A. Majumbar, and X. Xu, Nature {\bf 520},
  69   (2015).

\bibitem{li-CZNing-etal.17}
Y. Li, J. Zhang, D. Huang, H. Sun, F. Fan, J. Feng, Z. Wang, and C.~Z. Ning,
  Nature Nanotechnology {\bf 12},  987   (2017).

\end{thebibliography}

\end{document}